\documentclass[a4paper,conference]{IEEEtran} %[conference] %
\ifCLASSINFOpdf
  \usepackage[pdftex,dvipdfm]{graphicx}
\else
\fi
\usepackage[cmex10]{amsmath}
\usepackage{amssymb,verbatim}
\usepackage[dvipdfm]{graphicx}
\begin{document}
\title{%
An improvement of optical PPM communication with high security
}

\author{
\IEEEauthorblockN{Osamu HIROTA$^{1,2}$, Masaki SOHMA $^{1}$ \\}
\IEEEauthorblockA{
1. Quantum ICT Research Institute, Tamagawa University\\
6-1-1, Tamagawa-gakuen, Machida, Tokyo 194-8610, Japan\\
2. Research and Development Initiative, Chuo University, \\
1-13-27, Kasuga, Bunkyou-ku, Tokyo 112-8551, Japan\\
{\footnotesize\tt E-mail:hirota@lab.tamagawa.ac.jp, sohma@eng.tamagawa.ac.jp} \vspace*{-2.64ex}}
}

\maketitle

%%%%%%%%%%%%%%%%%%%%%%%%%%%%%%%%%%%%%%%%%%%%%%%%%%%%%%%%%%%%%%%%
%%%%%%%%%%%%%%%%%%%%%%%%%%%%%%%%%%%%%%%%%%%%%%%%%%%%%%%%%%%%%%%%
%%%%%%%%%%%%%%%%%%%%%%%%%%%%%%%%%%%%%%%%%%%%%%%%%%%%%%%%%%%%%%%%
%%%%%%%%%%%%%%%%%%%%%%%%%%%%%%%%%%%%%%%%%%%%%%%%%%%%%%%%%%%%%%%%
%%%%%%%%%%%%%%%%%%%%%%%%%%%%%%%%%%%%%%%%%%%%%%%%%%%%%%%%%%%%%%%%
\begin{abstract}
The purpose of this paper is to celebrate Sir D. Payne's 80th birthday by dedicating our latest results as a continuation of
 his work in optical communications.
One of the important issues in optical communications is to protect the transmission information data passing through optical fiber channels.
Many ideas have been proposed and are being actively developed to implement in the physical layer of ultra-high-speed communications. 
As a representative example,  a methodology that achieves this goal by fusing modulation techniques 
in ordinary optical communications and mathematical cipher is being actively studied. 
To further clarify this advantage, a theoretical concept has been proposed to solve the problem by adopting a PPM code format.
However, this scheme has several difficulties at the implementation stage. 
Thus, this paper presents an optical scheme on modulation and receiver that eliminate those drawbacks.

\end{abstract}
%%%%%%%%%%%%%%%%%%%%%%%%%%%%%%%%%%%%%%%%%%%%%%%%%%%%%%%%%%%%%%%%
%%%%%%%%%%%%%%%%%%%%%%%%%%%%%%%%%%%%%%%%%%%%%%%%%%%%%%%%%%%%%%%%
%%%%%%%%%%%%%%%%%%%%%%%%%%%%%%%%%%%%%%%%%%%%%%%%%%%%%%%%%%%%%%%%
%%%%%%%%%%%%%%%%%%%%%%%%%%%%%%%%%%%%%%%%%%%%%%%%%%%%%%%%%%%%%%%%
%%%%%%%%%%%%%%%%%%%%%%%%%%%%%%%%%%%%%%%%%%%%%%%%%%%%%%%%%%%%%%%%
%\pacs{02.30.Tb, 03.65.-w,03.65.Ta}

% \fi
\IEEEpeerreviewmaketitle

\section{Introduction}

The technologies that have had the social impact on communications technology in the 20th century are semiconductor lasers,
 fiber optics, and optical amplification. These technologies support the communication infrastructure of today's information society.
The most important feature of optical communication is its ability to transmit data over distances of thousands of kilometers 
at speeds in excess of 100 Gbps. 
It is well known that Sir D.Payne was a great leader of the team at Southampton that invented the erbium-doped fibre amplifier [1,2], 
and he pointed many optical communications researchers in the right direction.
 Furthermore, the device developments were carried out by his successors [3].
A number of researchers have stated that without contributions of D.Payne, 
the communication technology would not have developed into what it is today.

At the end of the 20th century, Northwestern University and M.I.T in the U.S. and Tamagawa University in Japan have begun to develop 
technologies to guarantee the security of communication infrastructures. But it is extremely difficult to develop cryptographic techniques
 to protect the information while maintaining the high speed and long-distance transmission performance by ``optical amplifiers".
In order to make clever use of the characteristics of optical communication, it is preferable to apply physical cipher, 
which is different from the conventional mathematical cipher. It is expected to provide a new research challenge for 
optical communication researchers.
One of candidates relates to a technology that applies quantum noise effects of light to enhance security by 
combining mathematical symmetric key cipher with optical modulation technology.
The basic concept of this type of technology is to prevent information leakage and tampering from the communication lines 
without sacrificing the performance of current optical communications.

To solve the above problem, H.P.Yuen established the basic concept of a new physical cipher under the collaboration of P.Kumar and O.Hirota, 
and he disclosed a physical example in a white paper of 2000 and in the QCMC of 2002 together with Kumar group [4].
His basic idea is based on the principle of hiding the ciphertext of mathematical cipher by quantum noise.
 Such a principle leads to a situation where the ciphertexts that can be 
received by receivers for legitimate and eavesdropper are different. 
In other words, it is an encryption technology that automatically converts the eavesdropper's signal (ciphertext) 
into noise using quantum noise of light when an eavesdropper who does not know the secret key receives a signal, 
thereby increasing the security of the communication data (see Fig.1). The back gorund is given in the appendix A-1.

A great deal of research and development has been conducted to implement this scheme.
Since they are composed of a fusion of optical modulation techniques and mathematical cipher, 
various modulation schemes in optical communications can be applied.
Examples include phase modulation [5], amplitude modulation (intensity modulation)[6], frequency modulation [7], 
and quadrature amplitude modulation [8].
In addition, several improvement studies have been done by many institutes [9,10,11,12,13,14,15,16].

These studies are called the standard models for the encription by quantum noise which corresponds to the first generation of this technology.
These mechanisms far surpass the security of conventional mathematical cipher and are practically sufficient. 
However, the pursuit of final goal of security in symmetric key cipher is necessary as a technological development.
It is to completely disable the eavesdropper's mathematical algorithm analysis and exhaustive search.
Therefore, research on improvements from standard systems to generalized scheme has been initiated (see appendix A-1). 
Two directions exist for it. One is to introduce artificial randomization techniques into the standard system and  
the other is to use coding techniques together like PPM [17].

With the aim of contributing to the latter, this paper introduces a method to improve 
the original PPM coding scheme with some difficulties. 
 It is based on a phase pulse position modulation consisted of multi frequencies. 
And we show a concrete scheme to implement the unitary transformation for a basic randomization based on 
the general unitary transformation theory for the Gaussian states [18,19].

\begin{figure}
\centering{\includegraphics[width=8cm]{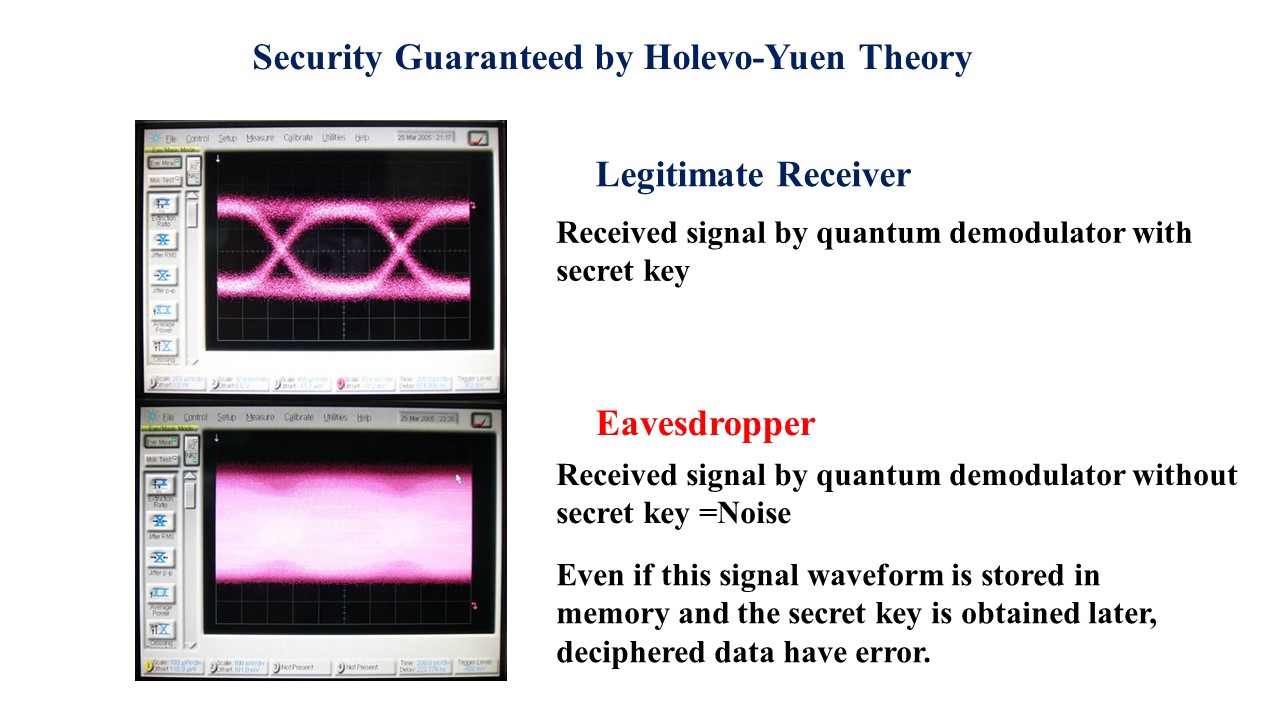}}
%\vskip 1mm
\caption{Conceptual eye pattern of quantun noise stream cipher. Upper side shows the decoded signal by Bob with secret. 
Lower side shows decoding result by Eve's receiver without secret key.}
%\vskip 2mm
\end{figure}

\section{Single frequency PPM scheme}
First, we describe an overview of the original scheme so called coherent PPM [17]. 
As a first step, a sequence of information bits $X =\{0,1\}$ is converted to $M$-valued signals. 
When a complex amplitude of a laser light pulse is $\alpha=|\alpha|\exp\{i\theta\}$, the quantum state vector of the light
 is in a coherent state $|\alpha >$, 
and when the signal is turned off, the light field is a vacuum and its quantum state vector is described as $|0>$. 
This is called the vacuum state and is a special state of the coherent state.
If we create a combination of amplitudes $\alpha:(\theta=0)$ and 0 with an optical pulse composed of a time length $t$ as one slot, 
the pulse train automatically has a quantum nature that is represented by a sequence of quantum state vectors of
 the combination of $|\alpha >$ and $|0>$.
Then, the $M$-valued data information corresponds to $M$-ary pulse position modulation ($M \gg 2$ ) as follows:
\begin{eqnarray}
S_1: |\Psi(X_1) > &=&|\alpha>|0 > |0 > \dots |0 > \nonumber \\
S_2 :|\Psi(X_2) >&=&|0 >|\alpha > |0 > \dots |0 > \nonumber \\
&&\vdots  \\
S_M:|\Psi(X_M) > &=&|0 >|0 > |0 > \dots |\alpha > \nonumber
\end{eqnarray}
These are to be regarded as a code word set $\{S_m\}, m=1,2,\dots, M$.

Next, when a $2M$-dimensional unitary transformation is applied to an optical pulse signal train
 with the above quantum state vector sequence of Eq.(1), the amplitude $\alpha:(\theta=0)$ localized at one place can be 
distributed to all slots. Then, the quantum state vector sequence of the pulse train automatically becomes
 one of the following signal set. 
\begin{eqnarray}
S_1^T:|\Phi_1 > &=&|\alpha_{11}>|\alpha_{21}> |\alpha_{31} > \dots |\alpha_{M1} > \nonumber \\
S_2^T:|\Phi_2 > &=&|\alpha_{12}>|\alpha_{22} > |\alpha_{32}> \dots |\alpha_{M2}> \nonumber \\
&&\vdots \\
S_M^T:|\Phi_M > &=&|\alpha_{1M}>|\alpha_{2M} >|\alpha_{3M}> \dots |\alpha_{MM} > \nonumber
\end{eqnarray}
where $\{\alpha_{l,m}\}$ is complex amplitude.
These $\{S^T_m\}$ are called random code word.
Such transformations are controlled by a running key sequence from  a pseudo-random number generator (PRNG) with a short secret key $K$.
Thus, one of the above set is randomly sent out into a communication channel as a ciphertext.

Here, let ${\bf U}_K=\{U(k^R_m)\}$, $m =1,2,3, \dots, M$ be the set of unitary transformation for this PPM modulation signal sequence.
The output runing key sequence from the pseudo-random number generator with a secret key $K$ 
is divided into $log M$-valued running key sequence $k^R$.
The running key sequence determines which unitary transformation is to be acted upon.
 When $U(k^R_m)$ is selected, if the input is from $|\Psi (X_1)>$ to $|\Psi (X_M)>$,
the resulting output quantum state sequences are as follows.
\begin{eqnarray}
|\Phi_1(k^R_m,X_1) >&=& U(k^R_m)|\Psi (X_1)>\nonumber \\
|\Phi_2(k^R_m,X_2) >&=&U(k^R_m)|\Psi (X_2)>\nonumber \\
&& \quad \quad \vdots \\
|\Phi_M(k^R_m,X_M) >&=& U(k^R_m)|\Psi (X_M)> \nonumber 
\end{eqnarray}

The regitemate receiver uses the information of the unitary transformation selected at the transmitter side with 
the same pseudo-random number as the sender to the optical signal with the random code word of Eq.(2).
That is, he transforms the random code word by the inverse unitary transformation to convert back to the original signal of Eq.(1).
As a result, he can discriminate them by photon counters almost without error at each slot as follows:
\begin{equation}
P(0|0)=1, \quad P(0|\alpha)=\exp \{-|\alpha|^2\}
\end{equation}
When $|\alpha |^2 \gg 1$, the average error probability for the code word is given as follows:
\begin{equation}
{\bar P^B_e}=\exp \{-|\alpha|^2\} \ll 1
\end{equation}

On the other hand, since the eavesdropper does not know which unitary transformation is used in the modulator, 
it is necessary to directly receive optical signals with the random code word of Eq.(3), and to identify the vast number of
 possible combinations of complex amplitudes. But, when $|\alpha|^2 \gg1$, the eavesdropper's error
 becomes ${\bar P}_e^E \longrightarrow 1$ only when $M\gg1$  [17].

Thus, the principle of the security is the difference between the error probabilities for the regitemate receiver and an eavesdropper
 tapping into the communication channel.
That is, the average error probability for code words upon reception of a signal system with quantum state vector sequences of
 Eq.(1) and Eq.(3) is overwhelmingly large for the latter, while that for the former is almost zero.

However, difficulties of two issues remain when trying to realize them. \\

(a) One is that it is difficult to realize a device that performs unitary transformations in the on-off based PPM, 
 because it involves the vacuum state $|0> $ in Eq.(1). This is one reason why no experimental attempts have existed to date.\\

(b) The other is that the length $M$ of Eq.(3) must be made considerably large to achieve the high security as denoted in the section IV.
It means that the delay of the encryption or letency becomes a major drawback.
When requiring error probabilities by the exponential speed of convergence and eliminating the delay characteristics, 
the following increase in the bandwidth is necessary.
\begin{equation}
W_{cppm} \approx M \times B_S
\end{equation}
where $M \rightarrow 2^{M^*}$. It means that when the speed of convergence of error probabilities and 
the associated delay effects are taken into account, one needs exponential increase of $M$ (Appendix A-1).
 The above quantity is extremely large as a baseband for receiver circuit.\\

In order to apply this technology to a practical optical communication system, 
we have to solve the above issues.

\section{Proposed multi frequency phase PPM}
To overcome the above issues (a) and (b), we adopt a generalization theory of the unitary transformation for PPM system [18,19]. 
Following the theory, we propose phase PPM scheme by multi frequencies.
Figure 2 shows a transmitter of the proposed scheme.
First, $M$ lasers with different frequencies are prepared in parallel as light sources.
Their quantum states are in coherent state, respectively.
\begin{equation}
|\alpha,\omega_1>,|\alpha,\omega_2>,\dots,|\alpha,\omega_M> 
\end{equation}
where $|\alpha,\omega_j>$ means a coherent state at each frequency.
Using these parallel light sources, a phase-pulse position modulation by phase $\theta=\pi$ corresponding to
 $M$ original plaintext $X_m$ is performed. So the pulse position with the phase of $\theta=\pi$ becomes information as follows:
\begin{equation}
S_1: |{\tilde \Psi} (X_1) > =
%\[
\left (
\begin{array}{c}
 |\alpha e^{i\pi}, \omega_1> \\
|\alpha,\omega_2> \\
\vdots \\
| \alpha,\omega_M>\\
\end{array}
\right )\nonumber
%\]
\end{equation}

\begin{equation}
S_2: |{\tilde \Psi} (X_2) > =
%\[
\left (
\begin{array}{c}
|\alpha, \omega_1> \\
|\alpha e^{i\pi},\omega_2> \\
 \vdots \\
|\alpha,\omega_M>\\
\end{array}
\right ) \nonumber
%\]
\end{equation}

\quad \quad \quad \quad  \quad $\vdots$

\begin{equation}
S_M: |{\tilde \Psi} (X_M) > =
%\[
\left (
\begin{array}{c}
|\alpha, \omega_1> \\
|\alpha,\omega_2> \\
 \vdots \\
|\alpha e^{i\pi},\omega_M>\\
\end{array}
\right ) 
%\]
\end{equation}
Here, as a feature of our PPM scheme by phase, we can apply the symplectic matrix theory as a realization of 
the unitary transformation ${\bf U}_K$ that performs phase randomization, and the symplectic matrix can be 
specified as follows [see appendix A-2].
\begin{eqnarray}
&&{\bf L}(k_R)=  \\
%\[
&& \left (
\begin{array}{ccccc}
e^{i\theta_{11}(k_R)}& 0 & \dots & 0 & 0\\
0 & e^{i\theta_{22}(k_R)}& 0 & \dots & 0\\
\vdots & \dots & \dots & \dots & \vdots \\
0 & \dots & \dots & 0 & e^{i\theta_{MM}(k_R)} \nonumber \\
\end{array}
\right )
%\]
\end{eqnarray}
where the phase value $\theta_{jj}$ of each frequency is selected from $J$-basis phases in Fig.3. 
But the relationship between phase signals and running key values is pre-determined.
Thus, the phases of output code word : $\{ \theta_{11}(k_R),\theta_{22}(k_R),\dots, \theta_{MM}(k_R) \}$ depend on the running key.
Here, we have adopted only the diagonal component of the unitary transformation for simplicity.
Since the number of type of unitary transformation is $J^M$, the block length $N_B$ of the running key increases, 
and the total number of the running key $|K^R_N|$ are as follows:
\begin{eqnarray}
N_B=\log_2 M &\longrightarrow&  M \log_2 J \nonumber \\
|K^R_N|=\frac{2^{|K|}}{\log_2 M} &\longrightarrow& \frac{2^{|K|}}{M\log_2 J}
\end{eqnarray}
where the length of the secret key  $|K|$ is more than 256. Thus, its effect is small.
In addition, the distance between the signals of the $J$-basis phases at each frequency is $2\pi|\alpha|/J$ is 
designed so that some signals enter the area of the quantum noise of the heterodyne receiver. It is called masking value by quantum noise.
To enhance the value, we require the following conditions.
\begin{equation}
2\pi|\alpha|/J < 1
\end{equation}
So it corresponds to $J >> 1, J > M$ for the large $|\alpha|$.

\begin{figure}
\centering{\includegraphics[width=8cm]{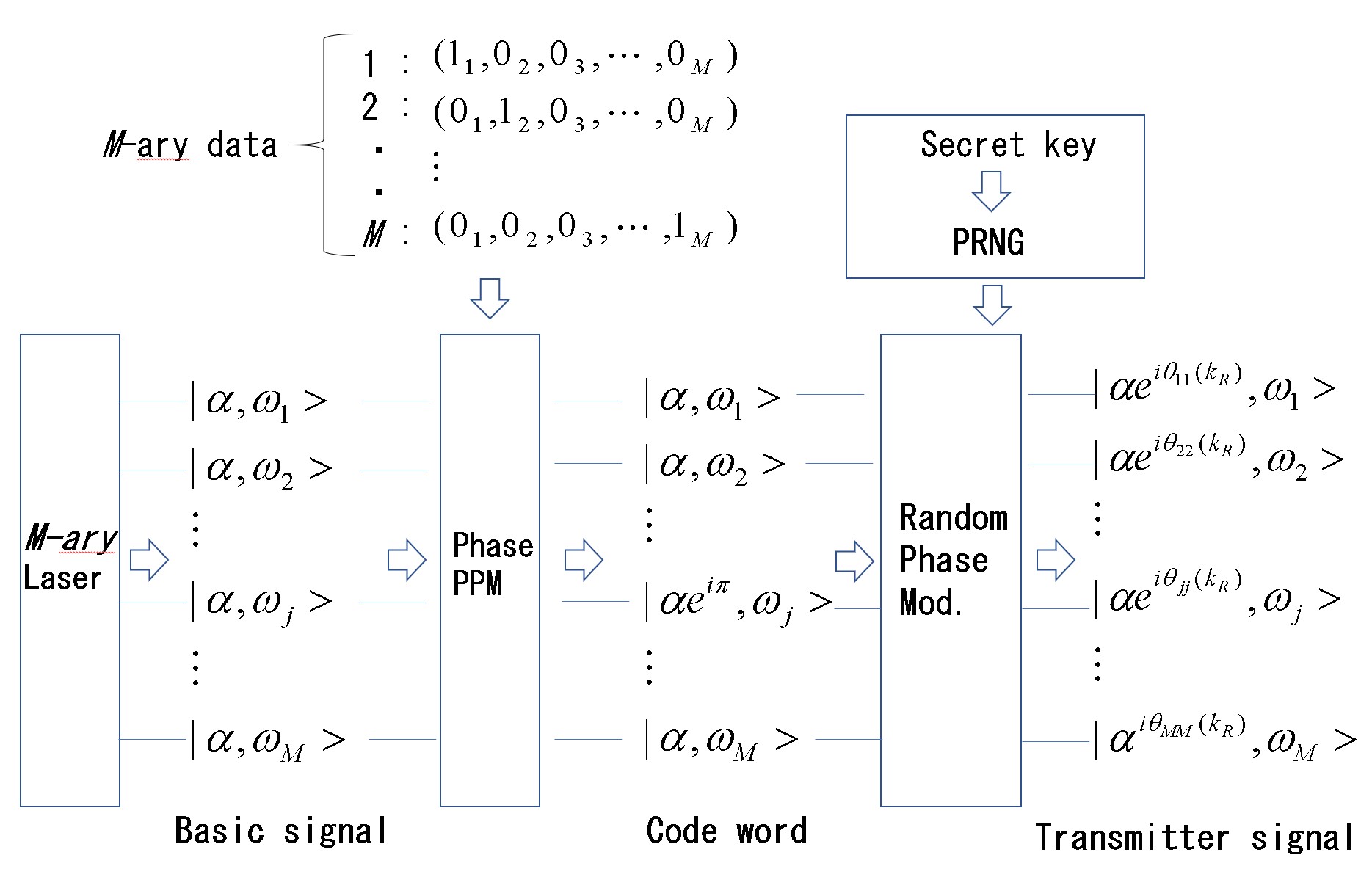}}
%\vskip 1mm
\caption{Transmitter scheme of proposed system. Phase PPM signals are randomly modulated by the matrix ${\bf L}(k_R)$ 
associated with unitary transformation ${\bf U}_K$ selected by running key from PRNG. }
%\vskip 2mm
\end{figure}

\begin{figure}
\centering{\includegraphics[width=5cm]{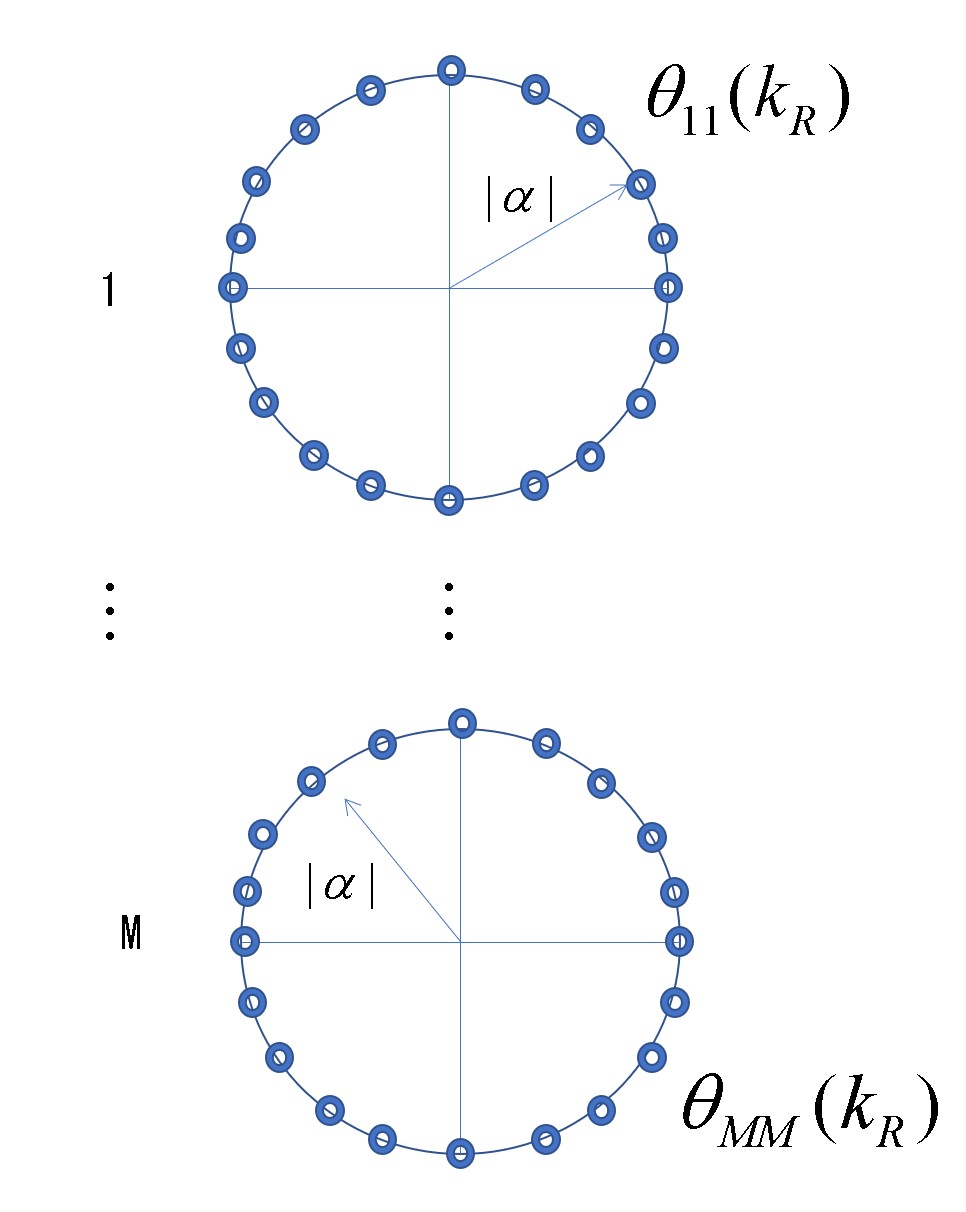}}
%\vskip 1mm
\caption{Optical phase signal constaration on phase space for each frequency. There are $J$ number of signals at each frequency. 
$M$ phases are simultaneously determined from $M$ frequency modes by the running key from PRNG. }
%\vskip 2mm
\end{figure}

Let us describe a receiver scheme.
The legitimate receiver acts on the random code words with the inverse transformation of the symplectic matrix of Eq.(9).
This action is controlled by the same running key as the sender.
As a result, the random code words  return to code word of the original phase multi frequency PPM.
So legitimate receiver can adopt the binary detection of phase 0 and $\pi$ for each frequency carrier wave by homodyne receivers, and 
he can determine the position of the phase  $\pi$ in $M$ frequencies with almost error free.

\section{Concrete system configuration}
\subsection{Randomization in transmitter}

The Figure 2 and Figure 3 show a transmitter system according to an embodiment, wherein the transmitter has a mode locked light source of $M$ frequencies. 
A pulse position modulation by phase at each frequency is created.
These are modulated by a binary phase shift modulator, and the phase PPM code words are generated.
$M$ pulses of different frequencies with optical signals expressed in code words are simultaneously input to a unitary transformation device.
The important issue in such a technology is how to realize the unitary transformation process to generate the random code. 
To solve this issue, we showed in the above section that one can adopt ``the symplectic matrix ${\bf L}$" to realize the randomization 
by unitary transformation. According to the theory, 
changing the phase of the complex amplitude of the coherent state can be achieved by changing the phase of the classical signal.
Thus, it can be realized with a classical multi-ary phase modulator.
The random code words are generated by such a phase modulator which corresponds to $2M$-dimensional unitary transformation
 on a quantum state vector sequence.
The running key sequence from PRNG controls the phase shift at each frequency to generate $J^M$ patterns of phase combination.
As shown in reference [18,19], such a transformation can produce an almost continuous waveform as its output. 
Then, the group of optical pulses of the random code word generated is transmitted over
 the communication channel as a ciphertext (random code word).
In the conventional scheme, the code word involves a vacuum state, making it difficult to realise unitary transformations in reality.
That is, under the constant total energy, an energy allocation operation is required between each signal.
So the phase PPM has an advantage in the real application.

\subsection{Receiver system for legitemate reception}

The Figure 4 shows the mechanism of the receiver (demodulator) of the legitimate receiver, when he or she receives random code words 
consisting of $M$ optical pulses of $M$ frequencies. 
They are inverted unitarily by a unitary transformation according to the same pseudo-random number as the sender. 
That is, the transformation is carried out, and the ciphertext with the random code word is converted back into
 a PPM signal with the original code word as shown in Eq.(8).
It can be done by the phase modulator that transforms the received random codes to the original code word.
The phase PPM signal at each frequency is identified by means of discrimination for 0 and 180 degrees of the phase of the pulse signal
at each frequency in the $M$ parallel homodyne receiver. The information data corresponds to the position of $\pi$ ($\theta$ =180) 
among $M$ frequencies.
We could adopt Helstrom optimal quantum measurement, but in reality it is difficult. So, if we were to adopt the usual homodyne receiver, 
the error probability would be as follows.
\begin{eqnarray}
&&{\bar P}_e^B \cong 1-\left \{erf(\frac{|\alpha|}{{\sqrt 2}\sigma_{ho}})\right \}^M \ll 1 \\
&&erf(y)= \frac{2}{\sqrt \pi}\int_0^{y} e^{-t^2} dt
\end{eqnarray}
where $|\alpha| \gg 1$, $\sigma_{ho}$ is the quantum noise effect in homodyne measurement given by
\begin{equation}
\sigma_{ho}= \frac{S}{4<n>_s}= \frac{|\alpha|^2}{4|\alpha|^2}=\frac{1}{4}
\end{equation}
$<n>_s$ is the shot noise. This error is small. If one requires the almost error free, it can be recovered with error correction code.

\begin{figure}
\centering{\includegraphics[width=8cm]{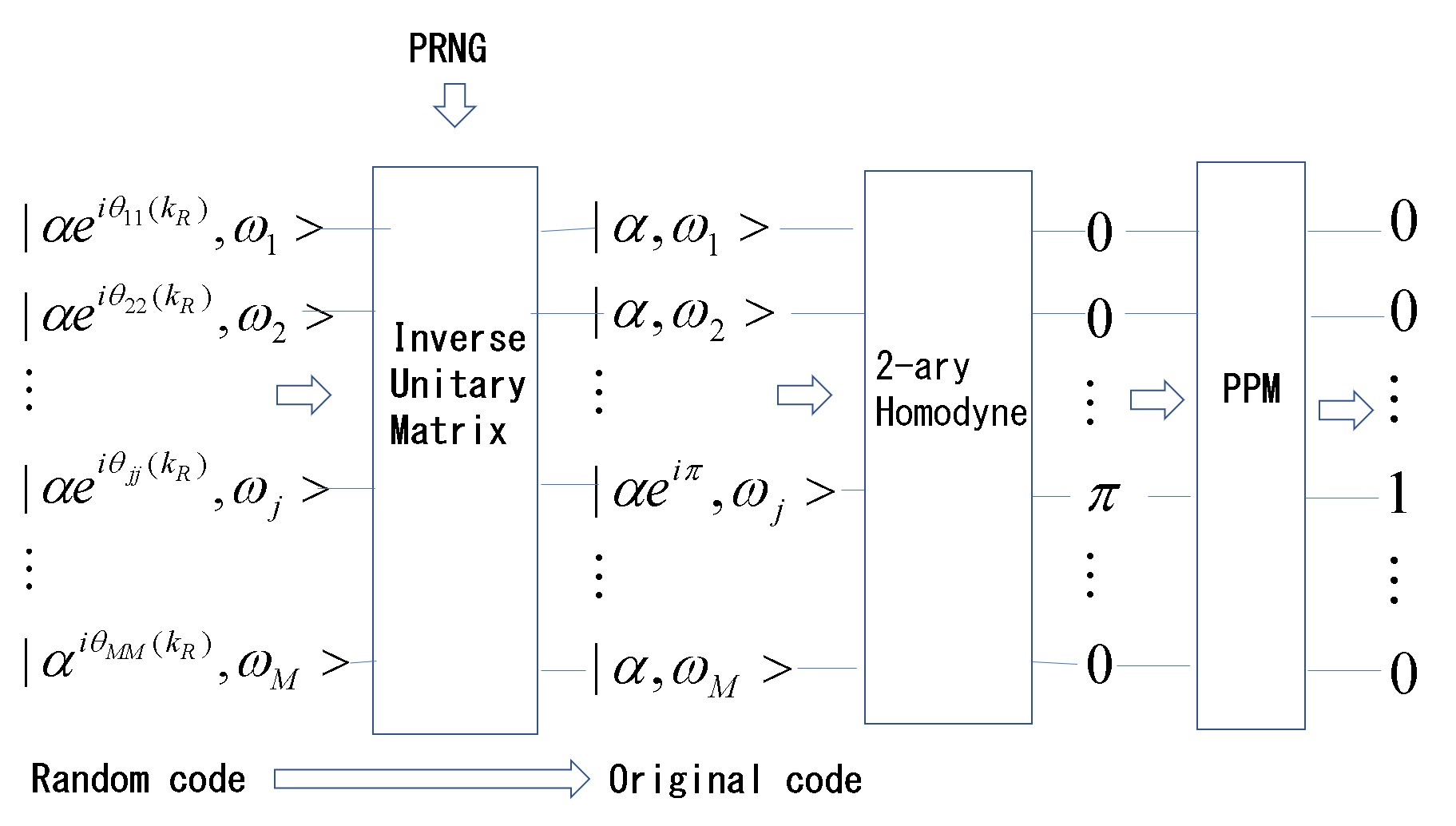}}
%\vskip 1mm
\caption{Legitemate receiver scheme of proposed system.The received signals are transformed into the original phase PPM. 
The binary detection is performed by homodayne receiver at each frequency. }
%\vskip 2mm
\end{figure}

\subsection{Receiver system for eavesdropper}

The Figure 5 shows the optimum receiver system for an eavesdropper when one adopts the proposed scheme.
Since the eavesdropper does not know the secret key (and pseudo-random numbers),
 she cannot carry out the inverse unitary transformation on the received random code words. 
Therefore, she has to estimate directly the optical signals as $M$-ary random code words. 
The general properties of quantum Bayes rule for the symmetric optical signals of multi parameter are given by Ban [20]. 
One of them is as follows:\\

${\bf Theorem 1}$ : 
If the signal set $\{|\alpha_j>\}, j=\{1,2, \dots, M\}$ is a symmetric, the optimum quantum measurement for each signal:
$\Pi_j$ is given by using Gram operator $H$ as follows:
\begin{eqnarray}
&&\Pi_j =|\mu_j><\mu_j|  \\
&&|\mu_j>= H^{-1/2}|\alpha_j> \nonumber \\
&&H= \sum_{j=1}^M |\alpha_j><\alpha_j| \nonumber
\end{eqnarray}
and the optimum  quantum Bayes solution is 
\begin{equation}
{\bar P}_e =1-|<\alpha_1|H^{-1/2}|\alpha_1>|^2
\end{equation}
\\
Let us assume that the eavesdropper adopts a simple decoding procedure with the above formula. 
 By applying calculation methods for the above formula on PSK of Osaki [21] and Kato [22] and 
also the quantum minimax theorem [23] to our scheme, 
we have the error probability for the eavesdropper in the phase PPM as follows: 
\begin{eqnarray}
{\bar P}_e^E &=&1- \left \{\frac{1}{J^2}(\sum_{l=1}^{J} {\sqrt \lambda_l})^2 \right \}^M  \\
\lambda_l&=&\sum_{k=1}^{J} <\alpha_1|\alpha_k >u^{-(k-1)l} \nonumber
\end{eqnarray}
where $u=\exp[i(2\pi/J)]$. This requres a numerical analysis for the visualization. Fortunately in the case of $J \gg 1$ in the Fig.3,
 a set of the quantum states can be treated as a quantum state system corresponding to a continuous signals. 
So the quantum optimum measurement can be approximated by Yuen-Lax quantum Cramer-Rao bound [24].\\

${\bf Theorem 2}$ : 
The estimation bound for complex amplitudes are given by following formula.
\begin{equation}
Var ({\hat \alpha}) \ge \frac{1}{Tr \rho {\cal L}{\cal L}^{\dagger} }
\end{equation}
where the right logarithm derivative is defined by 
\begin{equation}
\frac{\partial \rho}{\partial \alpha} = {\cal L}^{\dagger} \rho 
\end{equation}
And its solution is as follows:
\begin{equation}
{\cal L} = {\bf a}
\end{equation}
where ${\bf a}$ is a photon annihiration operator, and it corresponds to a heterodyne measurement.\\

As a result, one can approximate the optimum measurement in Eq.(15) by a heterodyne measurement.
Thus, to demonstrate intuitive understanding, we can assume that Eve adopts a heterodyne measurement.
Then we here can use the conventional Bayes rule, assuming  a priori probability is $1/J$ for each mode.
The error performance can be given from the application of Viterbi formula in the case of $J \gg 1$ and  $J^M \gg 1$ as follows:

The outputs of the heterodyne are the analog current denoted by $\{{\hat \theta}_{11}, {\hat \theta}_{22}, \dots, {\hat \theta}_{MM} \}$, 
in which each phase is estimated from complex amplitude.
The error probability of ciphertext in such a system based on PSK is given by 
\begin{equation}
{\bar P}_e^E \cong 1-\left \{erf(\frac{\Delta}{2\sigma_{he}})\right \}^M 
\end{equation}
where $\sigma_{he}=1$ is the quantum noise effect in heterodyne measurement.
$\Delta$ is the phase difference between neighboring phases in $J$-ary PSK in the Fig.3.
\begin{eqnarray}
&& \Delta=|\alpha|(1-\cos\delta) \\
&& \delta =2\pi/J
\end{eqnarray}
Thus, in the case of our scheme, one can control the error performance by $J$ under the fixed $M$ which is related to spectral efficiency.
So one can know that the error becomes as follows:
\begin{equation}
{\bar P}_e^E \longrightarrow 1, \quad J \gg 1, \quad M:fixed
\end{equation}

On the other hand, the lower bound of error probability of the conventional CPPM based on OOK by
random unitary transformation is given as follows [25]:
\begin{eqnarray}
{\bar P}_e^E &>& 1-\left \{\Phi(z)\right \}^n \Phi(z-2S) \\
 \Phi(z)&=& \frac{1}{\sqrt 2\pi}\int^z_{-\infty} exp\{-t^2/2\}dt \nonumber \\
n&=&\log M, \quad S=|\alpha|_{PPM}^2 \nonumber
\end{eqnarray}
Since $n=\log M$, $M$ must be made exponentially larger to obtain sufficient error.

Thus in our case, we do not need to make $M$ large in order to give a large error to an eavesdropper.
As a result, it is verified that the problem for exponential increase of bandwidth can be solved.

\begin{figure}
\centering{\includegraphics[width=8cm]{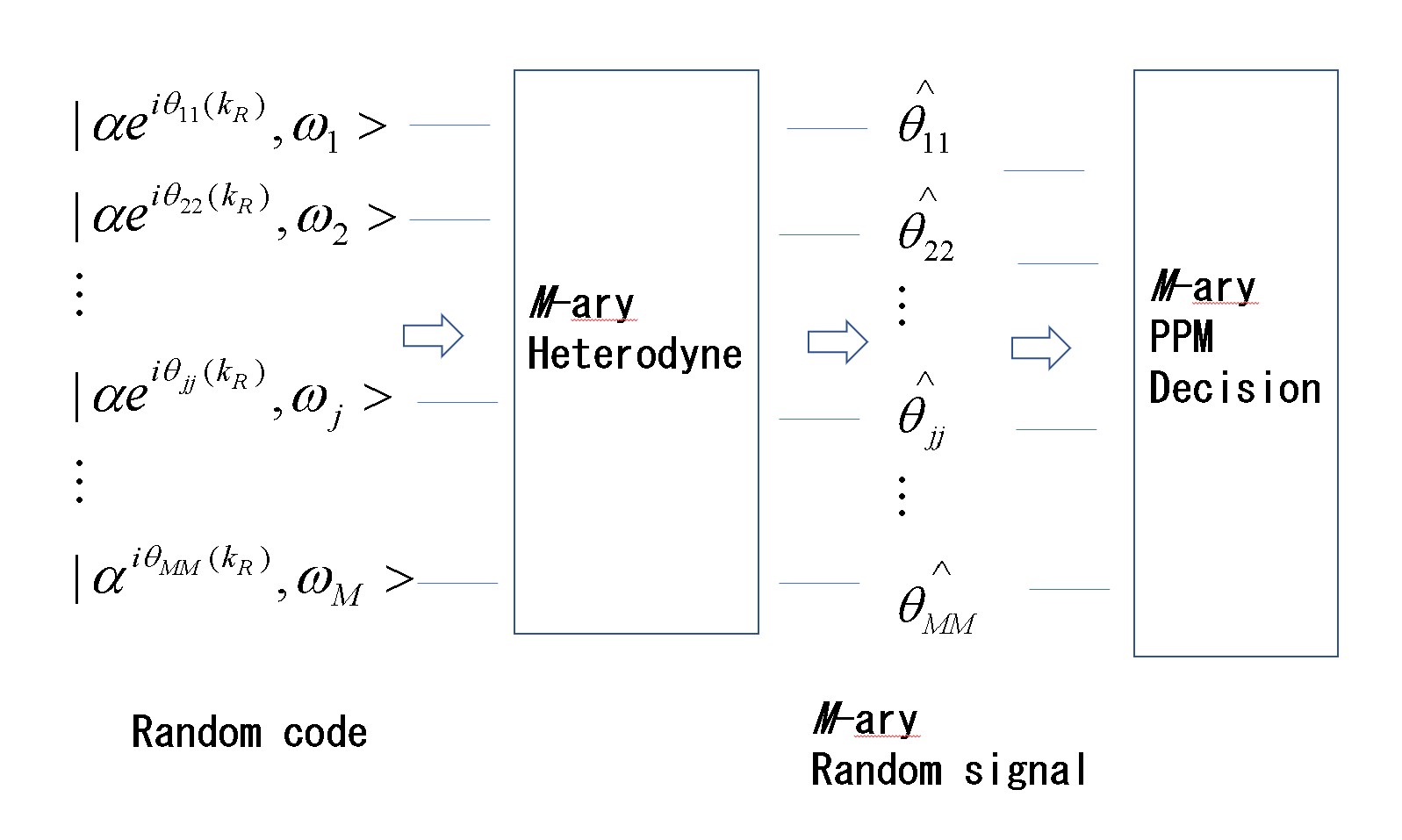}}
%\vskip 1mm
\caption{Eavesdropper's receiver scheme in proposed system. The random code signals are measured by
 $M$-ary heterodyne receivers consisted of $M$ 
frequency modes. From the output current, she estimates the PPM signals}
%\vskip 2mm
\end{figure}

\section{System performance}

The concept of CPPM [17] is a great idea to improve the security for quantum noise stream cipher.
 As we introduced in the introduction, 
however, there are two defects for the realization of such technologies.
One of them  is the realization of unitary transformation that can meet high-speed specifications 
and the other is the exponential increase in code length to increase security.
The latter requires an exponential increase in the baseband bandwidth of the communication system.

In the above sections, we have shown that the first issue can be solved by introducing a phase-PPM. On the other hand, 
the second problem can be solved by adopting a multi frequency phase-PPM scheme, which does not increase exponentialy like Eq(6) 
of the baseband bandwidth.
That is, the increase of the bandwidth for the fiber channel in the proposed scheme is given by 
\begin{equation}
W_{channel} =\Lambda \times B_{base} \ge 2B_{base}\times M
\end{equation}
where $B_{base}$ is the bandwidth of the baseband.

It is reasonable in the practical optical communication systems. This scheme can utilize a sufficiently large signal power.
In addition,  since optical fiber is extremely broadband, such an increase by sub-frequency is not a major technical drawback.
Our main challenge is to develop a multi-level phase modulator that operates at high speed. 
For the first demonstration, we will recommend $B_{base}\cong 1 GHz$, $M \cong 10$, $J \cong 6\times 10^4$ and $|\alpha|^2 \cong 10^3$.

\section{conclusion}

The basic concept of this technology is to protect optical communication channnels by utilizing ordinary optical communication technology.
 In particular, this technology controls the emergence phenomenon of quantum noise in light waves themselves by
 integrating optical modulation technology and mathematical cipher technology.
However, in general, the quantum noise is very small, and it is not sufficient to hide all signals.
To solve it, a PPM scheme has been proposed. But it has difficulties such as bandwidth expantion and realization of unitary transformation.
In this paper, we have proposed a scheme to improve the technical drawbacks of conventional PPM schemes, under the guidance of the reference [17].

Finally, we point out the fact that legitimate communicators can enjoy the benefits of normal optical communications.
In addition, since the reception performance between legitimate communicators and eavesdroppers is differentiated by
 the optimality in communication theory, quantitative evaluation also has the advantage of being presented
 more clearly than in conventional security technology.
We will report on detailed performance evaluations in the next paper and experimental studies of these schemes in the future papers.\\

\section*{Appendix}
\section*{A-1:Back ground of quantum stream cipher}

Here we present a back gound on the quantum noise stream cipher (QNSC).
The reference [17] and [25] provide a comprehensive overview of a new principles of quantum key distribution and direct encryption
 in which optical amplifiers can be adopted. However, the use of ordinary laser light limits the extent of quantum effects, 
and thus, realistic schemes that guarantee absolute security for key distribution remain unknown. 
On the other hand, the direct encryption relaxes the security conditions, 
and thus, Yuen-Kumar proposed a direct encryption scheme called $\alpha \eta$, which encrypts data as its first application. 

The security of conventional symmetric key ciphers based on the mathematics relays on the computational complexity.
 The purpose of the quantum stream cipher is to disable an eavesdropper's mathematical algorithm analysis 
and exhaustive search to symmetric key ciphers, which cannot be realized  by only the mathematical scheme.

The security of physical direct encryption for data such as CPPM is evaluated by the error probability of ciphertext to quantify 
the difficulty of algorithm analysis, which is based on application of Gallager theory [26] to Massey theory [27].
If the error probability of the ciphertext is one, it may be the dream security in the symmetric key cipher with PRNG.
 It means that one cannot apply any mathematical tool and exhaustive search, because the ciphertext becomes unknown [17][25][27][28].
Thus, the realization problem is so important.

\section*{A-2:General theory of symplectic transformation}
We show a theory for the general transformation of PPM consisted of coherent state by the unitary operator ${\bf U}$ associated with a symplectic transformation.
It will claim that any unitary operator composed of beamsplitters and phase shifters can be described by a symplectic transformation.

 First, let us consider a general $N-$ary coherent state as follows:
\begin{equation}
|\phi >=|\alpha_1>|\alpha_2> \dots |\alpha_N>
\end{equation}
The quantum characteristic function for the class of quantum Gaussian state is given as follows:
\begin{eqnarray}
\Phi ({\bf z})&=&Tr {\bf U}|\phi><\phi| {\bf U}^{\dagger} {\bf V}({\bf z}) \nonumber \\
&=&Tr |\phi><\phi|{\bf V}({\bf L}^T {\bf z}) 
\end{eqnarray}
where
\begin{eqnarray}
{\bf V}({\bf z})&=&\exp \{i {\bf R}^T {\bf z}\} \\
{\bf R}&=&[({\bf q}_1,{\bf p}_1), \dots ,({\bf q}_N,{\bf p}_N), ]^T 
\end{eqnarray}
and where $({\bf q}_i,{\bf p}_i)$ are the canonical conjugate operators. Then ${\bf L}$ is a symplectic matrix, and it is given by 
\begin{eqnarray}
%&&{\bf L}=  \\
%\[
&&{\bf  L}=\left (
\begin{array}{ccccc}
r_{11}e^{i\theta_{11}}& \dots & \dots &r_{1N}e^{i\theta_{1N}}   \\
r_{21}e^{i\theta_{21}}& \dots & \dots & r_{2N}e^{i\theta_{2N}}\\
\vdots & \dots & \dots & \vdots \\
r_{N1}e^{i\theta_{N1}} & \dots & \dots  & r_{NN}e^{i\theta_{NN}}  \\
\end{array}
\right )
%\]
\end{eqnarray}
Here let us denote a vector of complex amplitudes $\alpha$ as follows:
\begin{equation}
\vec{\alpha}_{in}=(\alpha_1, \alpha_2, \dots , \alpha_N)
\end{equation}
then we have the following relation.
\begin{equation}
\vec{\alpha}_{out}={\bf L} \vec{\alpha}_{in}=(\alpha_1^{out}, \alpha_2^{out}, \dots , \alpha_N^{out})
\end{equation}
As a result, the unitary transformation for the coherent state sequence is given by using Eq.(28)$\sim$ Eq.(33) as follows:
\begin{equation}
{\bf U}|\phi> =|\phi_{out}> =|\alpha_{1}^{out}>|\alpha_{2}^{out}> \dots |\alpha_{N}^{out}>
\end{equation}
Thus, Eq.(9) is a special case of the above formula. The more detailed theory is given in the references [18,19].\\

\end{document}